\documentclass[reprint,aps,prl]{revtex4-2}

\usepackage[pdftex]{graphicx}
\usepackage{amsmath}

\begin{document}

\title{Coherent phonon spectroscopy and interlayer modulation of charge density wave order in the kagome metal CsV$_3$Sb$_5$}

\author{Noah Ratcliff}
\affiliation{Materials Department, University of California, Santa Barbara, California 93106, USA}

\author{Lily Hallett}
\affiliation{Materials Department, University of California, Santa Barbara, California 93106, USA}

\author{Brenden R. Ortiz}
\affiliation{Materials Department, University of California, Santa Barbara, California 93106, USA}

\author{Stephen D. Wilson}
\affiliation{Materials Department, University of California, Santa Barbara, California 93106, USA}

\author{John W. Harter}
\email[Corresponding author: ]{harter@ucsb.edu}
\affiliation{Materials Department, University of California, Santa Barbara, California 93106, USA}

\date{\today}

\begin{abstract}
The recent discovery of the $A$V$_3$Sb$_5$ ($A=$ K, Rb, Cs) material family offers an exciting opportunity to investigate the interplay of correlations, topology, and superconductivity in kagome metals. The low energy physics of these materials is dominated by an unusual charge density wave phase, but little is understood about the true nature of the order parameter. In this work, we use a combination of ultrafast coherent phonon spectroscopy and first-principles density functional theory calculations to investigate the charge density wave order in CsV$_3$Sb$_5$. We find that the charge density wave is the result of a simultaneous condensation of three optical phonon modes at one $M$ and two $L$ points. This distortion can be described as tri-hexagonal ordering with an interlayer modulation along the $c$ axis. It breaks the $C_6$ rotational symmetry of the crystal and may offer a natural explanation for reports of uniaxial order at lower temperatures in this material family.
\end{abstract}

\maketitle

Identifying and characterizing novel phases of matter is at the forefront of materials science research. In this regard, kagome materials hold great promise for realizing exotic correlated and topological ground states~\cite{ko2009,guo2009,obrien2010,wen2010,yu2012,wang2013,kiesel2013,mazin2014,ye2018,kang2020}. It therefore comes as no surprise that the recently discovered quasi-two-dimensional vanadium-based kagome metal series $A$V$_3$Sb$_5$ ($A=$ K, Rb, Cs) has generated a flurry of interest~\cite{ortiz2019, ortiz2020, ortiz2021, yin2021, kenney2021, yang2020, yuX, jiangX, zhaoXa, chenXa, zhangX, chenXc, wangXa, liangX, tanX, liX, zhaoXb, chenXb, niX, uykurX, ortizX}. These materials possess a rare and coveted combination of nontrivial band topology and superconductivity. Furthermore, the high quality and exfoliable nature of single crystal samples makes them accessible to a wide variety of experimental techniques, which have already uncovered a complex web of intertwined properties. For example, the observation of a giant anomalous Hall effect~\cite{yang2020,yuX} and chiral charge order~\cite{jiangX} suggest proximity to a time-reversal symmetry breaking instability, even while there is no evidence of local moment magnetism~\cite{ortiz2020,kenney2021}. In addition, thermodynamic signatures of nodal quasiparticles~\cite{zhaoXa}, multiple superconducting domes~\cite{chenXa,zhangX,chenXc}, spin-triplet supercurrents~\cite{wangXa}, and zero-bias conductance peaks inside superconducting vortex cores~\cite{liangX} hint at the likelihood of unconventional---possibly topological---superconductivity.

The low energy physics of $A$V$_3$Sb$_5$ is dominated by an unusual charge density wave (CDW) phase transition at $T_\mathrm{CDW}=$ 78, 102, and 94 K for $A=$ K, Rb, and Cs, respectively~\cite{ortiz2020,ortiz2021,yin2021}. This phase transition can be clearly identified through anomalies in heat capacity, electrical resistivity, and magnetic susceptibility, and is widely suspected to be related to a Peierls-like nesting-driven instability at wave vectors connecting the $M$ points at the Brillouin zone boundary. However, the true nature of the CDW order parameter is not yet fully understood, and there are a number of open questions. One question concerns whether or not the CDW modulates along the $c$~axis~\cite{liangX,tanX,liX}. A second question involves reports of uniaxial CDW order which onsets well below $T_\mathrm{CDW}$~\cite{zhaoXb,liangX,chenXb,niX}. As superconductivity competes with and emerges from the CDW phase, understanding the CDW order in $A$V$_3$Sb$_5$ is essential. In this work, we use ultrafast optical transient reflectivity experiments in conjunction with first-principles density functional theory (DFT) calculations to study the CDW phase transition in CsV$_3$Sb$_5$. We show that the CDW is modulated along the $c$~axis and corresponds to a simultaneous condensation of three optical phonon modes ($3Q$ order) at one $M$ and two $L$ points. This ``MLL'' distortion may offer an explanation for the uniaxial order observed in CsV$_3$Sb$_5$ at lower temperatures.

\begin{figure*}[ht]
\includegraphics{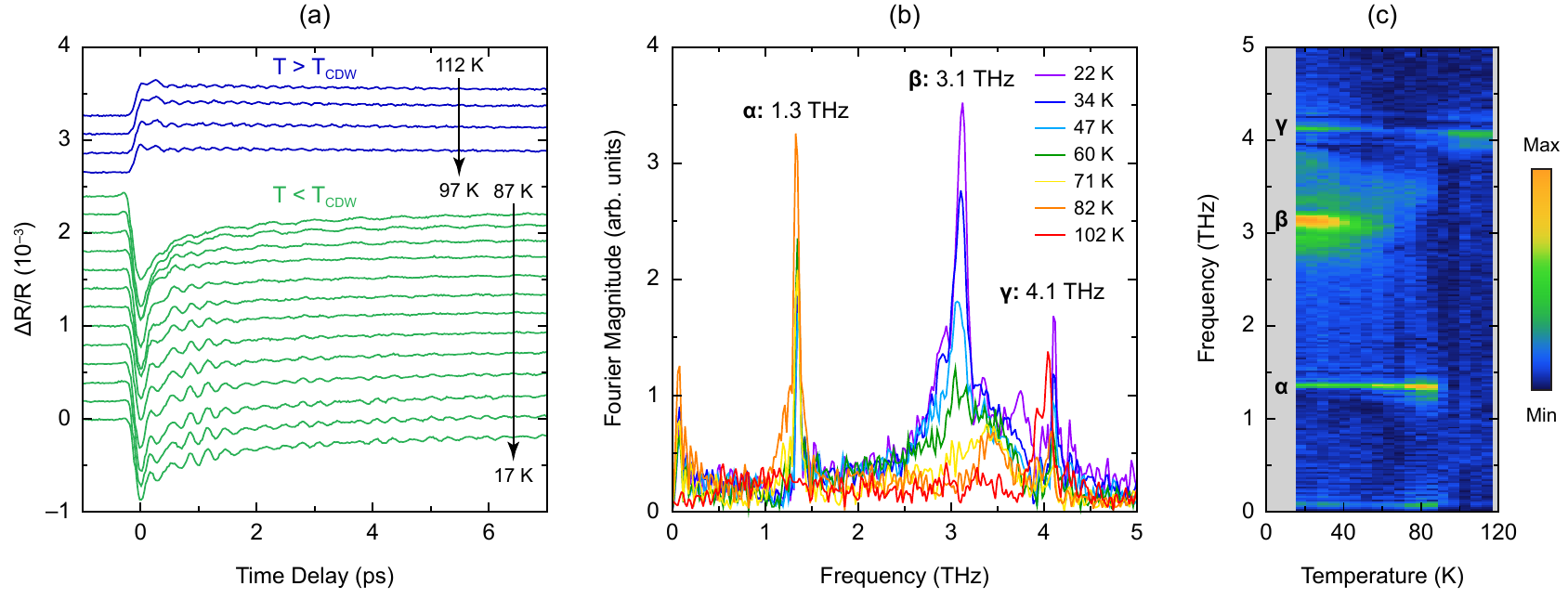}
\caption{\label{Fig1} Coherent phonon spectroscopy data for CsV$_3$Sb$_5$. (a)~Transient reflectivity curves for a series of temperatures above and below the CDW phase transition. Temperatures are approximately evenly spaced and scans are offset for clarity. Above $T_\mathrm{CDW}$, the reflectivity increases after the pump pulse and a single oscillation frequency is apparent. Below $T_\mathrm{CDW}$, in contrast, the reflectivity decreases after the pump pulse and the presence of multiple oscillation frequencies results in a complex beat pattern. (b)~Magnitude of the Fourier transform of the reflectivity oscillations after subtraction of a double exponential background. Three resonances are identified and labeled $\alpha$~(1.3~THz), $\beta$~(3.1~THz), and $\gamma$~(4.1~THz). (c)~Two-dimensional temperature-frequency map of the Fourier magnitude of the coherent phonon oscillations. While resonance $\gamma$ is present at all temperatures, resonance $\alpha$ only becomes Raman active below $T_\mathrm{CDW}$. Below $T^*\approx60$~K, the broad $\beta$ resonance appears and gradually grows in amplitude. A weak softening of all three frequencies is apparent with increasing temperature.}
\end{figure*}

Time-resolved optical reflectivity measurements were performed on freshly-cleaved surfaces of CsV$_3$Sb$_5$ single crystals mounted in an optical cryostat. Synthesis is described in Ref.~\citenum{ortiz2020}. A non-collinear optical parametric amplifier was used to generate $\sim50$~fs signal~(800~nm) and idler~(1515~nm) pulses at a repetition rate of 500~kHz, which were used as probe and pump beams, respectively. To avoid sample heating, both pulses had a low fluence of $\sim100$~$\mu$J/cm$^2$ and were linearly polarized in-plane. A lock-in amplifier and optical chopper were used to measure the small pump-induced transient change in reflectivity. Phonon frequencies were calculated using DFT within the Perdew-Burke-Ernzerhof generalized gradient approximation, as implemented in the Vienna ab initio Simulation Package (\textsc{vasp})~\cite{kresse1996}. The projector augmented wave potentials considered 9 valence electrons for the cesium atoms, the plane wave basis cutoff energy was 300~eV, and the zero damping DFT-D3 van der Waals correction was employed. The unit cell structure was relaxed using a $\Gamma$-centered $18\times18\times12$ k-point mesh. The relaxed lattice parameters were $a = b = 5.450$~\AA~and $c = 9.297$~\AA~and the out-of-plane antimony atoms were located at fractional height $z = 0.7435$. Phonon dispersion relations were calculated with the \textsc{phonopy} software package~\cite{togo2015} via the finite displacement method using a $3\times3\times2$ supercell. A $2\times2\times2$ supercell was used to calculate the energies of modes simultaneously condensing at the $M$ and $L$ points.

Figure~\ref{Fig1}(a) shows raw transient reflectivity data measured for CsV$_3$Sb$_5$ at several temperatures. Above $T_\mathrm{CDW}$, the reflectivity increases after the pump pulse and shows clear phonon oscillations at a single frequency. Below $T_\mathrm{CDW}$, however, the reflectivity decreases and exhibits a complex beat pattern indicative of the presence of multiple oscillation frequencies. The stark difference in the transient optical response across the CDW phase transition ($\Delta R/R > 0$ above $T_\mathrm{CDW}$, $\Delta R/R < 0$ below $T_\mathrm{CDW}$) is likely the result of changes in the density of states and the opening of a partial energy gap near the Fermi level at the CDW phase transition~\cite{uykurX}. Indeed, the sign of $\Delta R/R$ is known to depend sensitively on several electronic parameters influencing carrier dynamics, such as band filling and band gap renormalization~\cite{wells2014}; qualitative changes in the transient optical response are therefore unsurprising at $T_\mathrm{CDW}$.

To better understand the nature of the coherent phonon oscillations, a double exponential background ($A_0 + A_1e^{-t/\tau_1} + A_2e^{-t/\tau_2}$) is fitted and then subtracted from the reflectivity curves after $t = 100$~fs. The oscillations that remain are Fourier transformed and displayed in Fig.~\ref{Fig1}(b). Three resonance modes are clearly present in the material: $\alpha$ at 1.3~THz, $\beta$ at 3.1~THz, and $\gamma$ at 4.1~THz. No other frequencies are detected up to 10~THz. The striking temperature dependence of the coherent phonon oscillations can be more clearly seen in Fig.~\ref{Fig1}(c), which displays a two-dimensional temperature-frequency map of the oscillation spectrum. While the $\gamma$ mode is present at all temperatures, the $\alpha$ mode shows a sudden, intense appearance at $T \approx 92$~K. This temperature is close to $T_\mathrm{CDW} = 94$~K, as determined through independent heat capacity measurements~\cite{ortiz2020}, and we therefore identify it with the CDW critical temperature. The slightly lower detection temperature can be explained through modest local heating of the sample by the laser. At even lower temperatures, below $T^*\approx60$~K, the broad $\beta$ mode appears. All three resonances show a weak frequency softening as temperature is increased, but there is no evidence of complete softening ($\omega \rightarrow 0$) for any of the modes.

\begin{figure}[ht]
\includegraphics{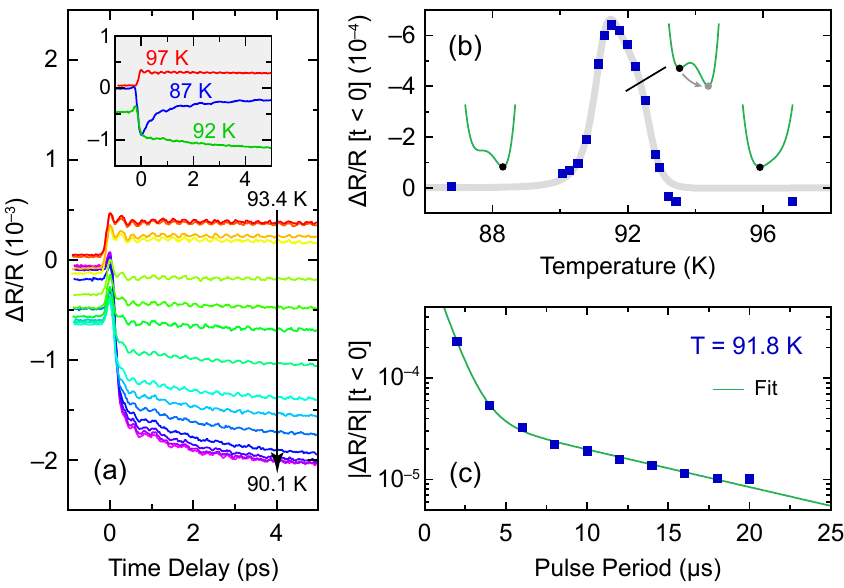}
\caption{\label{Fig2} Behavior near the CDW phase transition. (a)~Transient reflectivity curves for a finely-spaced series of temperatures across the CDW phase transition. Temperatures are approximately evenly spaced. Curves are \textit{not} offset and finite $\Delta R/R$ values for $t < 0$ are real. Inset shows pronounced changes in the transient optical response above (red), at (green), and below (blue) $T_\mathrm{CDW}$. (b)~Average $\Delta R/R$ value for time delays between $-1$~ps and $-0.2$~ps versus temperature. A metastable-like divergence in the lifetime of the transient optical response is observed at $T_\mathrm{CDW}$. Insets are cartoons of the free energy and the gray curve is a guide to the eye. (c)~Average $\left|\Delta R/R\right|$ value for negative time delays versus pump-probe pulse period at the CDW phase transition. The fit is described in the main text.}
\end{figure}

To investigate the critical behavior of the phase transition, transient reflectivity data were collected within a narrow temperature range centered at $T_\mathrm{CDW}$. As Fig.~\ref{Fig2}(a) illustrates, we observed a pronounced qualitative change in the shape of the transient reflectivity curve within $\sim1$~K of $T_\mathrm{CDW}$. In particular, we measured an anomalously long recovery behavior together with a finite $\Delta R/R$ value for \textit{negative} time delays, as shown in Fig.~\ref{Fig2}(b). This is likely the result of a pump-induced metastable change in the material, such that recovery to equilibrium after one pulse is still occurring by the time the next pulse arrives. To test this hypothesis, we measured the negative time delay anomaly as a function of the pulse period~$T$. Figure~\ref{Fig2}(c) shows that a double exponential recovery, $\Delta R(t<0)/R = A_1e^{-T/\tau_1} + A_2e^{-T/\tau_2}$, is consistent with our data, with a least squares fit giving $\tau_1 = 0.92$~$\mu$s and $\tau_2 = 11.7$~$\mu$s. These values represent extremely long lifetimes for typical electronic and phononic degrees of freedom in a solid, and instead point to metastable phase coexistence associated with a first-order phase transition. We therefore conclude that the CDW phase transition in CsV$_3$Sb$_5$ is discontinuous.

\begin{figure*}[ht]
\includegraphics{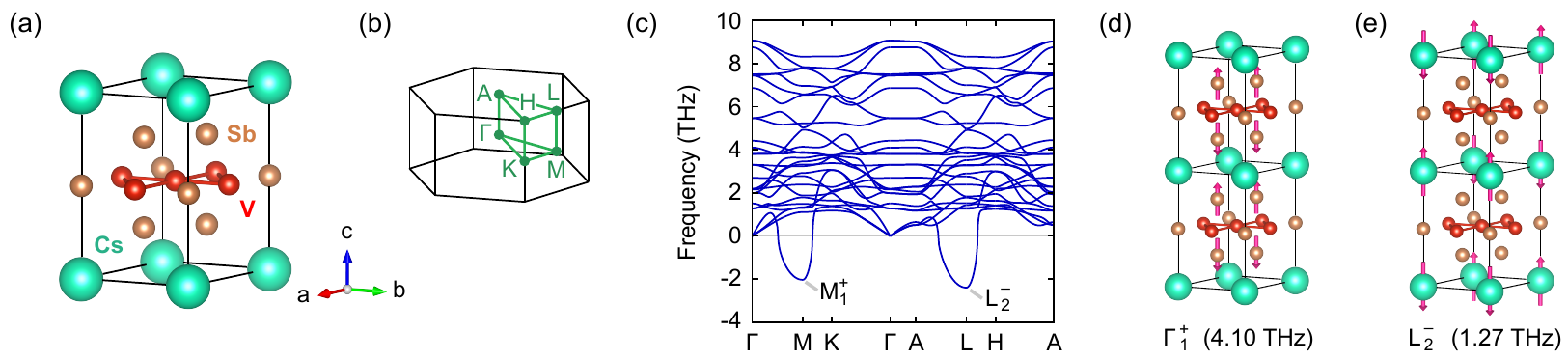}
\caption{\label{Fig3} First-principles phonon calculations. (a)~Unit cell for CsV$_3$Sb$_5$. The vanadium atoms (red) form a perfect kagome net and are coordinated with in-plane and out-of-plane antimony atoms. (b)~First Brillouin zone of the hexagonal lattice, with high-symmetry points labeled. (c)~Calculated phonon dispersion relations. There are two unstable modes with imaginary frequencies: one at the $M$ point with irreducible representation $M_1^+$, and the other at the $L$ point with irreducible representation $L_2^-$. (d)~Illustration of the fully symmetric $\Gamma_1^+$ phonon at 4.10 THz experimentally detected at all temperatures. (e)~Illustration of the $L_2^-$ phonon at 1.27 THz experimentally detected below $T_\mathrm{CDW}$.}
\end{figure*}

\begin{table}[t]
\caption{\label{Table1} Calculated phonon frequencies up to 5 THz.}
\begin{ruledtabular}
\begin{tabular}{cccccc}
\multicolumn{2}{c}{$\Gamma$ point} & \multicolumn{2}{c}{$M$ point} & \multicolumn{2}{c}{$L$ point} \\
\cline{1-2} \cline{3-4} \cline{5-6}
Irrep & $f$ (THz) & Irrep & $f$ (THz) & Irrep & $f$ (THz) \\
\hline
$\Gamma_6^-$ & 1.28 & $M_1^+$ & $-2.03$ & $L_2^-$ & $-2.40$ \\
$\Gamma_2^-$ & 1.51 & $M_2^-$ &  1.11 & $L_2^-$ &  1.27 \\
$\Gamma_6^-$ & 1.99 & $M_3^-$ &  1.30 & $L_3^-$ &  1.30 \\
$\Gamma_4^-$ & 2.16 & $M_3^+$ &  1.33 & $L_4^-$ &  1.43 \\
$\Gamma_6^+$ & 2.19 & $M_4^-$ &  1.49 & $L_4^-$ &  1.52 \\
$\Gamma_2^-$ & 2.70 & $M_2^+$ &  1.79 & $L_1^-$ &  1.78 \\
$\Gamma_5^-$ & 3.30 & $M_2^-$ &  1.88 & $L_1^+$ &  1.84 \\
$\Gamma_5^+$ & 3.81 & $M_1^-$ &  2.39 & $L_2^+$ &  2.39 \\
$\Gamma_1^+$ & 4.10 & $M_2^-$ &  2.78 & $L_1^+$ &  2.76 \\
$\Gamma_3^+$ & 4.42 & $M_4^-$ &  2.92 & $L_3^+$ &  2.90 \\
             &      & $M_4^+$ &  3.24 & $L_3^-$ &  3.27 \\
             &      & $M_3^-$ &  3.32 & $L_4^+$ &  3.29 \\
             &      & $M_3^+$ &  3.46 & $L_4^-$ &  3.52 \\
             &      & $M_1^+$ &  3.63 & $L_2^-$ &  3.67 \\
             &      & $M_2^-$ &  3.85 & $L_1^+$ &  3.84 \\
             &      & $M_4^-$ &  3.89 & $L_3^+$ &  3.87 \\
             &      & $M_3^-$ &  4.14 & $L_4^+$ &  4.12 \\
             &      & $M_4^-$ &  4.93 & $L_3^+$ &  4.92 \\
             &      &         &       & $L_2^-$ &  4.98 \\
\end{tabular}
\end{ruledtabular}
\end{table}

\begin{figure}[b]
\includegraphics{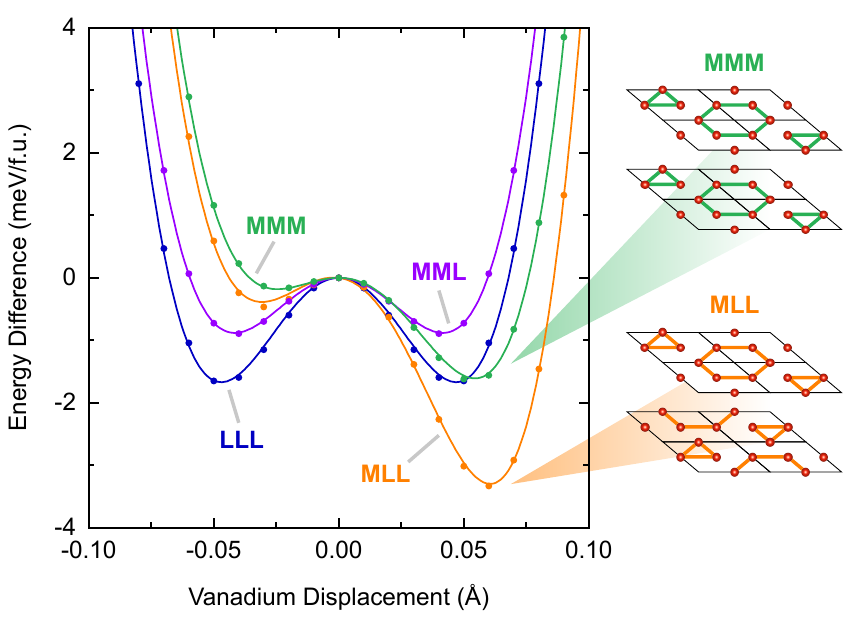}
\caption{\label{Fig4} Energy lowering by $3Q$ distortions. For MMM and MLL configurations, positive displacements correspond to tri-hexagonal distortions and negative displacements correspond to ``Star of David'' distortions. The lowest energy occurs for the MLL configuration consisting of in-plane tri-hexagonal distortions that are laterally shifted in neighboring planes. Curves are least squares fits to sixth-degree polynomials.}
\end{figure}

To explain our observations, we performed DFT calculations of the phonon mode frequencies in CsV$_3$Sb$_5$. At high temperatures, the material crystallizes in the $P6/mmm$ space group, as shown in Fig.~\ref{Fig3}(a). Of particular note in the first Brillouin zone is the $M$ point with $k_z=0$, which connects saddle points in the kagome electronic band structure and is associated with an ostensible Peierls-like nesting-driven instability~\cite{ortiz2020}, and the $L$ point with $k_z=\pi/c$, as shown in Fig.~\ref{Fig3}(b). Table~\ref{Table1} lists all calculated $\Gamma$-, $M$-, and $L$-point phonon frequencies and irreducible representations up to 5~THz, and Fig.~\ref{Fig3}(c) shows the full phonon dispersion relation. Two unstable modes are predicted, one at the $M$ point and one at the $L$ point, with the latter having a slightly larger imaginary frequency. A symmetry analysis indicates that the irreducible representations of the modes are $M_1^+$and $L_2^-$, respectively, using the notation of Miller and Love~\cite{millerlove}.

As shown previously, a simultaneous condensation of the three equivalent $M$-point modes leads to a fully symmetric breathing mode with two possible configurations: a ``Star of David'' distortion (breathing out), or a tri-hexagonal---also called ``inverse Star of David''---distortion (breathing in)~\cite{tanX}. These two distortions, which we label MMM, are only one possible type of $3Q$ order, however. Other possible combinations of $M$- and $L$-point modes exist, including MML, MLL, and LLL. To determine which of these four possibilities represents the most likely ground state in CsV$_3$Sb$_5$, we calculated their energies as a function of atomic displacement. For these calculations, we assume a frozen phonon model and do not further relax the unit cell (i.e. nonlinear phonon couplings are ignored). As Fig.~\ref{Fig4} shows, the lowest energy state corresponds to the tri-hexagonal MLL distortion and not the MMM distortion as widely believed. This ground state distortion, consisting of one $M$- and two $L$-point modes, can be described as in-plane tri-hexagonal distortions with a lateral shift of one lattice vector between neighboring kagome planes. Further support for this conclusion comes from our earlier experimental determination that the CDW phase transition is first-order, which immediately eliminates the odd-parity MML and LLL distortions.

Now that the true CDW structure of CsV$_3$Sb$_5$ has been determined, we can revisit our coherent phonon spectroscopy data. Firstly, the $\gamma$ mode at 4.1~THz, which is detected at all temperatures, must be a fully symmetric mode. Sure enough, the DFT calculations find only a single $\Gamma_1^+$ ($A_{1g}$) mode which exactly matches the experimental frequency. This mode, illustrated in Fig.~\ref{Fig3}(d), consists of coherent motion of the out-of-plane antimony atoms along the $c$~axis towards/away from the kagome planes. The excellent agreement between experiment and theory confirms the accuracy of our DFT calculations. We next address the $\alpha$ mode detected at 1.3~THz, which appears abruptly at $T_\mathrm{CDW}$ and therefore represents a phonon that becomes Raman active in the CDW phase. This indicates that the $\alpha$ mode must have either $M_1^+$ or $L_2^-$ symmetry. Indeed, there is an $L_2^-$ phonon with a frequency of 1.27~THz, which is illustrated in Fig.~\ref{Fig3}(e) and involves cesium atom motion along the $c$~axis. The observation of this mode therefore serves as strong experimental confirmation that the CDW order in CsV$_3$Sb$_5$ is of the MLL type and modulated along the $c$~axis. We also note that an $M_1^+$ phonon is predicted at 3.63~THz and a second $L_2^-$ phonon is predicted at 3.67~THz. Weak spectral intensity is discernible in the Fourier map below $T_\mathrm{CDW}$ near these frequencies, suggesting that these modes also make an appearance in the data, albeit with a weaker amplitude.

Finally, we examine the $\beta$ resonance, which extends over a broad spectral range centered at 3.1~THz. This is an unusual spectral feature because it seems to appear at $T^*\approx60$~K, well below $T_\mathrm{CDW}$. For this reason, we believe this resonance is related to the uniaxial order observed below this temperature by scanning tunneling microscopy experiments~\cite{zhaoXb,liangX,chenXb}. A $1Q$ uniaxial ordering would break the $C_6$ rotational symmetry of the crystal and result in a myriad of newly Raman-active modes excitable by a pump pulse, several with frequencies near 3.1~THz. In fact, MLL order \textit{already} breaks the $C_6$ rotational symmetry (whereas MMM order does not), and we speculate that this may offer a natural explanation for the uniaxial order reported in this material system. We further speculate that the lower onset temperature $T^*$ of the $1Q$ order may be related to an order-disorder crossover associated with $c$-axis coherence, as there are three equivalent MLL configurations which are distinguished only by the direction of the lateral plane-to-plane shift of the tri-hexagonal CDW pattern. While these ideas appear promising, further experimental and theoretical work will be needed to fully solve the $1Q$ puzzle in CsV$_3$Sb$_5$.

In conclusion, we used coherent phonon spectroscopy experiments in conjunction with first-principles DFT calculations to investigate the unusual CDW order in CsV$_3$Sb$_5$. We uncovered a first-order $3Q$ phase transition characterized by a simultaneous condensation of one $M$- and two $L$-point phonons, which we call ``MLL'' CDW order. This ordering, distinct from the more commonly discussed ``MMM'' ordering, involves interlayer modulation of the CDW along the $c$ axis and may offer a natural explanation for the uniaxial order observed at lower temperatures. We have only considered $k_z=\pi/c$ ($L$ point) wave vectors in this work, but it is possible that the true ground state of the system could have a longer wavelength modulation, as recently suggested by x-ray diffraction experiments~\cite{ortizX}. A $4c$ wavelength, for example, could arise from a chiral screw-axis CDW in which the plane-to-plane shift direction continuously rotates. While many puzzles still surround the $A$V$_3$Sb$_5$ materials, our results represent a large step forward in understanding the CDW order in CsV$_3$Sb$_5$.


We would like to thank Sam Teicher and Hengxin Tan for helpful discussions. This work was supported by the National Science Foundation (NSF) through Enabling Quantum Leap: Convergent Accelerated Discovery Foundries for Quantum Materials Science, Engineering, and Information (Q-AMASE-i): Quantum Foundry at UC Santa Barbara (DMR-1906325). Use was made of computational facilities purchased with funds from the NSF (CNS-1725797) and administered by the Center for Scientific Computing (CSC). The CSC is supported by the California NanoSystems Institute and the Materials Research Science and Engineering Center (MRSEC; NSF DMR-1720256) at UC Santa Barbara. L.H. acknowledges support from the Roy T. Eddleman Center for Quantum Innovation at UC Santa Barbara. B.R.O. acknowledges support from the California NanoSystems Institute through the Elings Fellowship program.

\end{document}